\documentclass{ws-procs961x669}            
\begin{document}
\title{Cosmology in the novel scalar-tensor representation of $f(R,T)$ gravity}

\author{Tiago B. Gon\c{c}alves}

\address{Instituto de Astrof\'{i}sica e Ci\^{e}ncias do Espa\c{c}o, Faculdade de Ci\^{e}ncias da Universidade de Lisboa,\\ Edif\'{i}cio C8, Campo Grande, PT1749-016 Lisbon, Portugal\\
E-mail: tgoncalves@alunos.fc.ul.pt}

\author{Jo\~{a}o Lu\'{i}s Rosa}

\address{Institute of Physics, University of Tartu, W. Ostwaldi 1,\\
50411 Tartu, Estonia\\
E-mail: joaoluis92@gmail.com}

\author{Francisco S. N. Lobo}

\address{Instituto de Astrof\'{i}sica e Ci\^{e}ncias do Espa\c{c}o, Faculdade de Ci\^{e}ncias da Universidade de Lisboa, Edif\'{i}cio C8, Campo Grande, PT1749-016 Lisbon, Portugal\\
E-mail: fslobo@fc.ul.pt}

\begin{abstract}
We apply cosmological reconstruction methods to the $f(R,T)$ modified gravity, in its recently developed
scalar-tensor representation. We do this analysis assuming a perfect fluid in a Friedmann-Lema\^{i}tre-Robsertson-Walker (FLRW) universe. In this contribution we show the equations of motion obtained and we present the solutions found for one of the particular cases we analysed: an exponential
evolution of the cosmological scale factor. 
\end{abstract}


\bodymatter
\section{Introduction}
The late-time universe is observed to be expanding in an accelerated way \cite{SupernovaCosmologyProject:1998vns,SupernovaSearchTeam:1998fmf}. This behaviour can be modelled by introducing a dark energy component, often identified as a cosmological constant. However, we still lack the understanding for the dark energy mechanism at the fundamental level. Alternatively, it is possible to explain the accelerated expansion by modifying the theory of gravity. Many such modifications have been proposed to extend Einstein's General Relativity (GR) (e.g. see  Refs. \citenum{Nojiri:2006ri,Lobo:2008sg,Nojiri:2010wj,Clifton:2011jh,Capozziello:2011et,CANTATA:2021ktz,Avelino:2016lpj}). For instance, one can modify the gravitational Lagrangian, which depends linearly on the Ricci curvature scalar $R$, to instead depend on a \emph{general function} of $R$ -- thus, referred to as $f(R)$ gravity\cite{Sotiriou:2008rp}. Resulting in modified Friedmann equations, $f(R)$ gravity can allow for the accelerated expansion without the need for dark energy\cite{Capozziello:2002rd}.

It is natural, then, to explore further extensions, such as $f(R,T)$ gravity\cite{Harko:2011kv} where the action depends on a general function of the curvature scalar $R$ \emph{and} of the trace of the stress-energy tensor $T$. Because of this explicit coupling between curvature and matter, the stress-energy tensor is not necessarily conserved, allowing the appearance of an extra force. Since its proposal, the $f(R,T)$ gravity has sparked a lot of interest leading, in some cases, to the study of its cosmological applications (e.g. see Refs. \citenum{Jamil:2011ptc,Houndjo:2011fb,Houndjo:2011tu,Alvarenga:2013syu,Shabani:2013djy,Shabani:2014xvi,Moraes:2015kka}).

Recently, the original geometrical representation of the $f(R,T)$ gravity was reformulated into an equivalent scalar-tensor representation\cite{Rosa:2021teg}, where the two extra degrees of freedom of the theory are expressed by means of two scalar fields. So far, this scalar-tensor theory was used to find thin shells' junction conditions  \cite{Rosa:2021teg} and thick brane solutions\cite{Rosa:2021tei}. Since in scalar-tensor theories it is often easier to find analytical solutions, it calls for further study also of cosmological models in this representation of the $f(R,T)$ gravity. This study is undertaken in our work, modelling a Friedmann-Lema\^{i}tre-Robsertson-Walker (FLRW) universe. We used reconstruction methods\cite{Capozziello:2005ku,Multamaki:2005zs,Jamil:2011ptc,Houndjo:2011fb,Houndjo:2011tu}, whereby the function $f(R,T)$ is not defined {\it a priori\/}, and instead we start from known cosmological solutions such as the time evolution of the spatial scale factor which seem to model well the observations. Here we give simply a preview of this work and some of its preliminary results. A more complete analysis is investigated in Ref. \citenum{Goncalves:2021vci}.

\section{The theory of the $f(R,T)$ gravity}

We first introduce briefly the theory of $f(R,T)$ gravity, showing the main equations without detailing all of the steps. For further details we point to the main references. In its original geometrical formulation (see Ref. \citenum{Harko:2011kv}), the action of the $f(R,T)$ gravity is
\begin{equation}\label{eq:fRTaction-original}
S = \frac{1}{2\kappa^2} \int\sqrt{-g} \, f(R,T) d^4 x+ \int \sqrt{-g} \, \mathcal{L}_m d^4 x,
\end{equation}
where $\kappa^2=8\pi G/c^4$, $G$ is the gravitational constant and $c$ is the speed of light, $g$ is the determinant of the metric $g_{\mu\nu}$, $f\left(R,T\right)$ is an arbitrary well-behaved function of the Ricci curvature scalar $R$ and the trace of the stress-energy tensor $T$, and $\mathcal{L}_m$ is the matter Lagrangian. More explicitly, $R=g^{\mu\nu}R_{\mu\nu}$, where $R_{\mu\nu}$ is the Ricci tensor, and $T=g^{\mu\nu}T_{\mu\nu}$, where the stress-energy tensor $T_{\mu\nu}$ is defined in terms of the variation of the matter Lagrangian $\mathcal L_m$ as
\begin{equation}
T_{\mu\nu}=-\frac{2}{\sqrt{-g}}\frac{\delta\left(\sqrt{-g}\mathcal L_m\right)}{\delta g^{\mu\nu}}.
\end{equation}

When reformulating into the dynamically equivalent scalar-tensor representation (see Ref. \citenum{Rosa:2021teg} for further details), one defines two scalar fields
 \begin{equation}\label{eq:varphi&psi}
\varphi\equiv\frac{\partial f}{\partial R} ,\qquad
\psi\equiv\frac{\partial f}{\partial T},
\end{equation}
with the associated potential
\begin{equation}\label{eq:potential}
V(\varphi,\psi) \equiv -f(R,T)+ \varphi R + \psi T .
\end{equation}
With these definitions, the action \eqref{eq:fRTaction-original} can equivalently be written as
\begin{equation}\label{eq:STaction}
S = \frac{1}{2\kappa^2} \int_{\Omega} \sqrt{-g} \left[\varphi R+\psi T - V(\varphi, \psi)\right]d^4 x 
+ \int_{\Omega} \sqrt{-g} \mathcal{L}_m d^4 x . 
\end{equation}

The variation of this action with respect to the metric $g_{\mu\nu}$ yields the modified field equations
\begin{equation}\label{eq:fields}
\varphi R_{\mu\nu}-\frac{1}{2}g_{\mu\nu}\left(\varphi R + \psi T - V\right)-(\nabla_\mu\nabla_\nu-g_{\mu\nu}\square)\varphi = \kappa^2 T_{\mu\nu} -\psi (T_{\mu\nu} + \Theta_{\mu\nu}),
\end{equation}
where $\nabla_\mu$ is the covariant derivative and $\square\equiv\nabla^\sigma\nabla_\sigma$ is the D’Alembert operator, both defined in terms of the metric $g_{\mu\nu}$, and the tensor $\Theta_{\mu\nu}$ is defined as
\begin{equation}\label{eq:Theta-varT}
\Theta_{\mu\nu}\equiv g^{\rho\sigma}\frac{\delta T_{\rho\sigma}}{\delta g^{\mu\nu}}.
\end{equation}
On the other hand, the variation of the action \eqref{eq:STaction} with respect to the scalar fields $\varphi$ and $\psi$ gives, respectively,
\begin{equation}\label{eq:Vphi}
V_{\varphi} = R,
\end{equation}
\begin{equation}\label{eq:Vpsi}
V_{\psi} = T,
\end{equation}
where the subscripts in $V_\varphi$ and $V_\psi$ denote the partial derivatives of the potential $V(\varphi,\psi)$ with respect to the variables $\varphi$ and $\psi$, respectively.

A general conservation equation can be found by taking the divergence of Eq.~\eqref{eq:fields}, 
\begin{equation}\label{eq:conserv-general2}
(\kappa^2-\psi)\nabla^\mu T_{\mu\nu}=\left(T_{\mu\nu}+\Theta_{\mu\nu}\right)\nabla^\mu\psi+\psi\nabla^\mu\Theta_{\mu\nu}-\frac{1}{2}g_{\mu\nu}\left[R\nabla^\mu\varphi+\nabla^\mu\left(\psi T-V\right)\right].
\end{equation}
Even though it is not a requirement of the theory, in this work we will impose the conservation of the stress-energy tensor, i.e. $\nabla^\mu T_{\mu\nu}=0$, a condition that is automatically assured in GR. As detailed in the following section, for our cosmological model we use a FLRW metric and a perfect fluid approach in order to find the modified equations of motion. 

In the following, we adopt a system of geometrized units in such a way that $G=c=1$, and thus $\kappa^2=8\pi$. 

\section{Cosmology in the scalar-tensor representation}
\subsection{Assumptions and equations of motion}
Let us first state the assumptions of our cosmological model with which we derived the equations of motion that are the focus of our study. We take the metric to be of the FLRW form, i.e., in spherical coordinates $(t,r,\theta,\phi)$,
\begin{equation}\label{eq:FLRW-metric}
ds^2 = -dt^2+a^2(t)\left[\frac{dr^2}{1-kr^2}+r^2\left(d\theta^2+\sin^2\theta d\phi^2\right)\right], 
\end{equation}
where $a(t)$ is the scale factor and $k$ is the curvature parameter which can take the values $k=\left\{-1,0,1\right\}$ corresponding to a hyperbolic, spatially flat, or hyperspherical universe, respectively.

We assume the stress-energy tensor $T_{\mu\nu}$ to be well described by that of an isotropic perfect fluid, 
\begin{equation}\label{eq:em-fluid}
T_{\mu\nu}=(\rho+p)u_\mu u_\nu +p g_{\mu\nu},
\end{equation}
where $\rho$ is the energy density, $p$ is the isotropic pressure, and $u^\mu$ is the fluid 4-velocity satisfying the normalization condition $u_\mu u^\mu=-1$. We take the matter Lagrangian to be $\mathcal L_m=p \hspace{5pt}$ \cite{Bertolami:2008ab}. Under these considerations, the tensor $\Theta_{\mu\nu}$ takes the form
\begin{equation}\label{eq:Theta-fluid}
\Theta_{\mu\nu}=-2T_{\mu\nu}+p g_{\mu\nu}.
\end{equation}

With these assumptions, and with all physical quantities depending only on the time coordinate $t$ (which guarantees the homogeneity and isotropy of the solutions $\rho\left(t\right)$, $p\left(t\right)$, $\varphi\left(t\right)$, $\psi\left(t\right)$, etc.), the following equations of motion were obtained. From the two independent components of Eq.~\eqref{eq:fields}, we obtained the modified Friedmann equation and the modified Raychaudhuri equation,
\begin{equation}\label{eq:tt}
\dot{\varphi}\left(\frac{\dot{a}}{a}\right) + \varphi \left( \frac{\dot{a}^{2} + k}{a^{2}} \right)  = \frac{8 \pi}{3}\rho + \frac{\psi}{2}\left( \rho - \frac{1}{3}p\right)+ \frac{1}{6} V,
\end{equation}
\begin{equation}\label{eq:rr}
\ddot{\varphi}+ 2\dot{\varphi}\left(\frac{\dot{a}}{a}\right) +\varphi\left( \frac{2\ddot{a}}{a} + \frac{\dot{a}^{2}+k}{a^{2}} \right)  = -8\pi p  + \frac{\psi}{2}\left(\rho-3p\right) + \frac{1}{2} V,
\end{equation}
respectively, where overdots denote derivatives with respect to time. From Eqs.~\eqref{eq:Vphi} and \eqref{eq:Vpsi} we obtained 
\begin{equation}\label{eq:Vphicosmo}
V_{\varphi} = 6\left( \frac{\ddot{a}}{a}+ \frac{\dot{a}^{2}+k}{a^{2}}\right),
\end{equation}
\begin{equation}\label{eq:Vpsicosmo}
V_{\psi} = 3p-\rho,
\end{equation}
respectively. By our choice, we imposed $\nabla_\mu T^{\mu\nu} = 0$ that leads to the usual continuity equation,
\begin{equation}\label{eq:conserv-m}
\dot{\rho} = -3\frac{\dot{a}}{a}(\rho+p).
\end{equation}
Consequently, from the general conservation equation (Eq.~\eqref{eq:conserv-general2}) we obtained
\begin{equation}\label{eq:conserv-psi}
2\dot{\psi}(\rho+p) = -\psi(\dot{\rho}-\dot{p}).
\end{equation} 
Finally, we impose an equation of state of the form
\begin{equation}\label{eq:state}
p=w\rho,
\end{equation}
where $w$ is a dimensionless parameter. 

These equations of motion are the framework with which we searched for cosmological solutions in the $f(R,T)$ gravity. There is a redundancy in the equations obtained, since they are not all linearly independent, and one can be discarded without loss of generality. We chose to work with Eqs.~\eqref{eq:tt}, \eqref{eq:Vphicosmo}, \eqref{eq:Vpsicosmo}, \eqref{eq:conserv-m}, \eqref{eq:conserv-psi}, and \eqref{eq:state} which form a system of six independent equations. This system is still underdetermined since we have nine degrees of freedom: $a$, $k$, $w$, $\rho$, $p$, $\varphi$, $\psi$ and $V$ which contributes with two degrees of freedom (because it is a function of two variables, $\varphi$ and $\psi$). In a first instance, we solved as much as it was possible without further constraining the system, obtaining some general solutions for $\rho(t)$, $p(t)$, $\psi(t)$, and the $V(\varphi,\psi)$ dependence on $\psi$. Only subsequently, did we provide further constraints to close the system and analyse particular cases.

\subsection{General solutions}
Taking the system of equations we have derived, Eqs.~\eqref{eq:tt}, \eqref{eq:Vphicosmo}, \eqref{eq:Vpsicosmo}, \eqref{eq:conserv-m}, \eqref{eq:conserv-psi}, and \eqref{eq:state}, we first found some general results. Beginning in a similar way that we would in GR, we used the equation of state \eqref{eq:state} in the continuity equation \eqref{eq:conserv-m} to find the general forms of the energy density and pressure as a function of $a$,
\begin{equation}\label{eq:solrho}
\rho = \rho_0 \left(\frac{a}{a_0} \right)^{-3(1+w)}, 
\end{equation}
\begin{equation}\label{eq:solp}
p = w \rho_0 \left(\frac{a}{a_0} \right)^{-3(1+w)},
\end{equation}
respectively, where $\rho_0$ and $a_0$ are arbitrary constants from integration, and usually the subscript $0$ denotes the value of the quantity today, at $t=t_0$. These solutions are identical to those in GR. This is only so because we required $\nabla_\mu T^{\mu\nu} = 0$ which, again, does not have to be generally true in modified gravity. Yet, it seems sensible to assume the validity of the continuity equation, at least on a first analysis. 

Equipped with the results in Eqs.~\eqref{eq:solrho} and \eqref{eq:solp}, we then solved Eq.~\eqref{eq:conserv-psi} by integration and obtained
\begin{equation}\label{eq:solpsi}
\psi=\psi_0 \left(\frac{a}{a_0}\right)^{\frac{3}{2}\left(1-w\right)},
\end{equation}
where $\psi_0$ is an integration constant. At this point, we were able to make the following considerations. The partial derivative $V_\psi$ in Eq.~\eqref{eq:Vpsicosmo} depends on $\rho$ and $p$. We have the forms of $\rho(a)$ and $p(a)$ in Eqs.~\eqref{eq:solrho} and \eqref{eq:solp}, and we can find a function $a(\psi)$ by inverting Eq~\eqref{eq:solpsi} (which we can do for $w\neq1$, so, in what follows we will not be considering stiff matter). Thus, we combined these results to write the partial derivative $V_\psi$ in Eq.~\eqref{eq:Vpsicosmo} as depending only on $\psi$. We integrated the partial derivative to obtain a separable potential $V\left(\varphi,\psi\right)=V_0+V_1\left(\varphi\right)+V_2\left(\psi\right)$, where $V_0$ is an arbitrary constant, $V_1(\varphi)$ is an arbitrary function of $\varphi$, and $V_2(\psi)$ is a function of $\psi$ given by
\begin{equation}\label{eq:solV2_psi}
V_2(\psi)=\frac{(1-3w)(1-w)}{(1+3w)}\rho_0 \psi_0^{\frac{2(1+w)}{(1-w)}}\psi^{-\frac{(1+3w)}{(1-w)}},
\end{equation}
which is undefined when $w=\left\{-1/3,1\right\}$ and vanishes when $w=1/3$ (corresponding to the case when the stress-energy tensor is traceless $T=0$, such as for radiation).

This is as far as we were able to go without imposing further constraints. Up to this point, from the system of equations we are considering we have already used Eqs.~\eqref{eq:Vpsicosmo},~\eqref{eq:conserv-m},~\eqref{eq:conserv-psi},~and~\eqref{eq:state}, so we have only Eqs.~\eqref{eq:tt} and \eqref{eq:Vphicosmo} remaining . In balance, there are still five degrees of freedom at the moment: $a$, $k$, $w$, $\varphi$ and $V_1(\varphi)$. Hence, we have to impose three further constraints in order to fully determine the system. We note that, as it stands, making a choice of the curvature and equation of state parameters, $k$ and $w$, is not sufficient to determine the evolution of the universe modelled by $a(t)$. This is in contrast with GR where to each equation of state corresponds a unique evolution of the scale factor $a(t)$. Therefore, in this work we took the approach of first specifying the form of the scale factor, and then explored the parameter space of $k$ and $w$ in each particular case. The forms of the scale factor we chose are those known to model well different epochs of the universe: $a(t)$ given by an exponential or by a power law. By way of example, we present here only the particular case with an exponential scale factor, which can model the accelerated expansion observed in the present era. 

\subsection{The particular case of an exponential expansion}
The universe is seen to be undergoing a phase of accelerated expansion. So, one of the particular cases we analysed was the de-Sitter solution that one obtains in GR, i.e., we impose in our system of equations an exponentially evolving scale factor of the form
\begin{equation}\label{eq:a-dS}
a(t)=a_0 e^{\sqrt{\Lambda}\left(t-t_0\right)},
\end{equation}
where $a_0$, $t_0$ and $\Lambda$ are constants. Once again, this particular choice does not require a unique value of $w$ in $f(R,T)$ gravity and the solutions for $\rho$ and $\psi$ from Eqs.~\eqref{eq:solrho} and \eqref{eq:solpsi} become
\begin{equation}\label{eq:solrho-dS}
\rho(t) = \rho_0  e^{-3\sqrt{\Lambda}\left(t-t_0\right)\left(1+w\right)},
\end{equation}
\begin{equation}\label{eq:solpsi-dS}
\psi(t)=\psi_0  e^{\frac{3}{2}\sqrt{\Lambda}\left(t-t_0\right)\left(1-w\right)},
\end{equation}
respectively. The pressure is simply given by $p\left(t\right)=w\rho\left(t\right)$.

To make further progress in order to find analytical solutions for $\varphi(t)$ and $V_1(\varphi)$, we set $k=0$ (flat geometry) and left $w$ arbitrary. With this choice of the curvature parameter and with the scale factor given by Eq.~\eqref{eq:a-dS} the computation of Eq.~\eqref{eq:Vphicosmo} yields simply $V_\varphi=12\Lambda$. Given that this is equivalent to having $dV_1/d\varphi=12\Lambda$, we integrated it to obtain $V_1\left(\varphi\right)=12\Lambda\varphi$. By doing so, we absorbed the arbitrary integration constant into $V_0$ in the full expression of the potential, which in this case reads as
\begin{equation}\label{eq:solV-dS-k0}
V(\varphi,\psi) = V_0 + 12\Lambda \varphi+\frac{(1-3w)(1-w)}{(1+3w)}\rho_0 \psi_0^{\frac{2(1+w)}{(1-w)}}\psi^{-\frac{(1+3w)}{(1-w)}}.
\end{equation}	
Having determined the form of the potential, we solved Eq.~\eqref{eq:tt} to obtain $\varphi(t)$,
\begin{equation}\label{eq:solphi-dSk0}
\varphi(t)=\varphi_0 e^{\sqrt{\Lambda}\left(t-t_0\right)} -\frac{V_0}{6\Lambda} 
-\frac{\rho_0}{3\Lambda}\bigg[ \frac{8\pi e^{-3\left(1+w\right)\sqrt{\Lambda}\left(t-t_0\right)}}{\left(4+3w\right)}
+ \frac{4\left(1+w\right)\psi_0 e^{-\frac{3}{2}\left(1+3w\right)\sqrt{\Lambda}\left(t-t_0\right)}}{\left(1+3w\right)\left(5+9w\right)}\bigg],
\end{equation}
where $\varphi_0$ is an integration constant. This solution $\varphi\left(t\right)$ is undefined at $w=\left\{-4/3,-5/9,-1/3\right\}$, and, from before, we were already not considering the $w=1$ case. Apart from these discontinuities, the system does not pose further constraints to the equation of state. In the paper under preparation, we include a more complete analysis of this parameter space. However, here we show the solutions only for three particular choices of $w$. 

A natural choice was to take $w=-1$,  which in GR is the equation of state of constant dark energy density that drives an exponential expansion, corresponding to a cosmological constant. Given that we have the freedom to do so, we also set $w=0$, i.e. the case in which pressureless (or dust-like) matter dominates. Alternatively, we also set $w=1/3$ to model the case in which the energy density is dominated by radiation (remembering, for instance, that the stress-energy tensor of the electromagnetic field is traceless, i.e. $T=3p-\rho=0$ in the  case of isotropic perfect fluid considered, corresponding to $w=1/3$). That we tried different values for the equation of state, e.g. for matter and radiation, may be confusing at first given that we are considering an exponential growth of the universe, but we need only to remember that due to the extra degrees of freedom in $f(R,T)$ gravity, this cases can still be consistent within the system of equations we derived. 

\begin{figure}
	\centering
	\includegraphics[width=0.6\columnwidth]{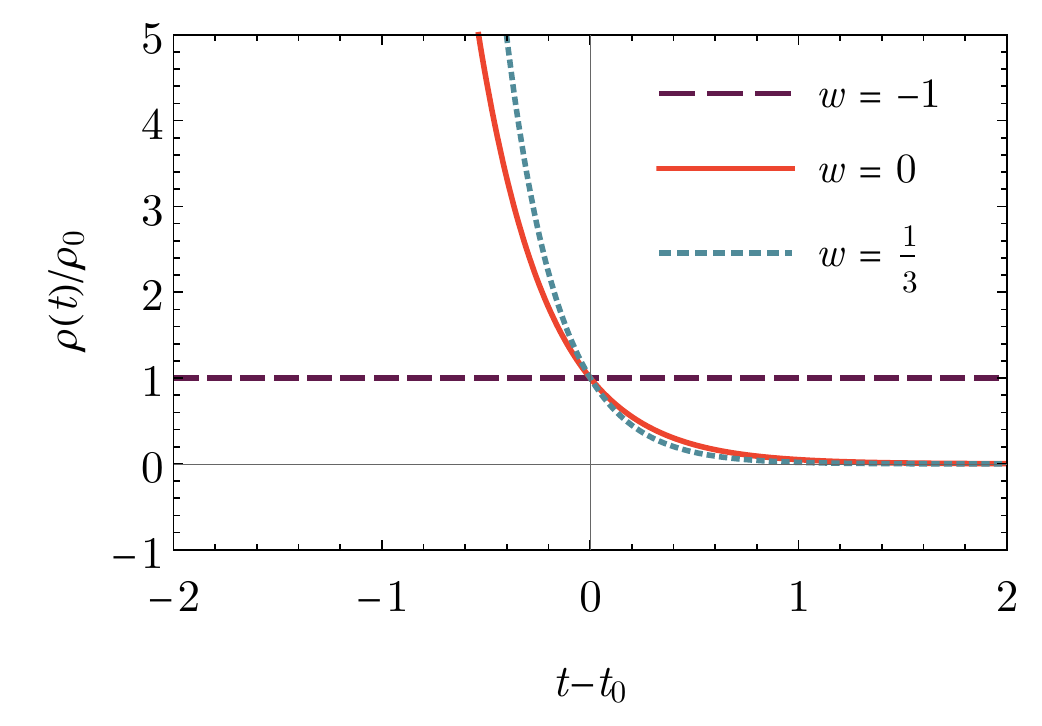}
	\caption{Energy density $\rho\left(t\right)/\rho_0$ from Eq.~\eqref{eq:solrho-dS} in the particular case where $a\left(t\right)$ is given by Eq.~\eqref{eq:a-dS}, setting $\Lambda=1$, with the equation of state forced to be each of these values $w=\{-1,0,1/3\}$.}
	\label{fig:Plot_dS_k0_rho}
\end{figure}

\begin{figure}
	\centering
	\includegraphics[width=0.6\columnwidth]{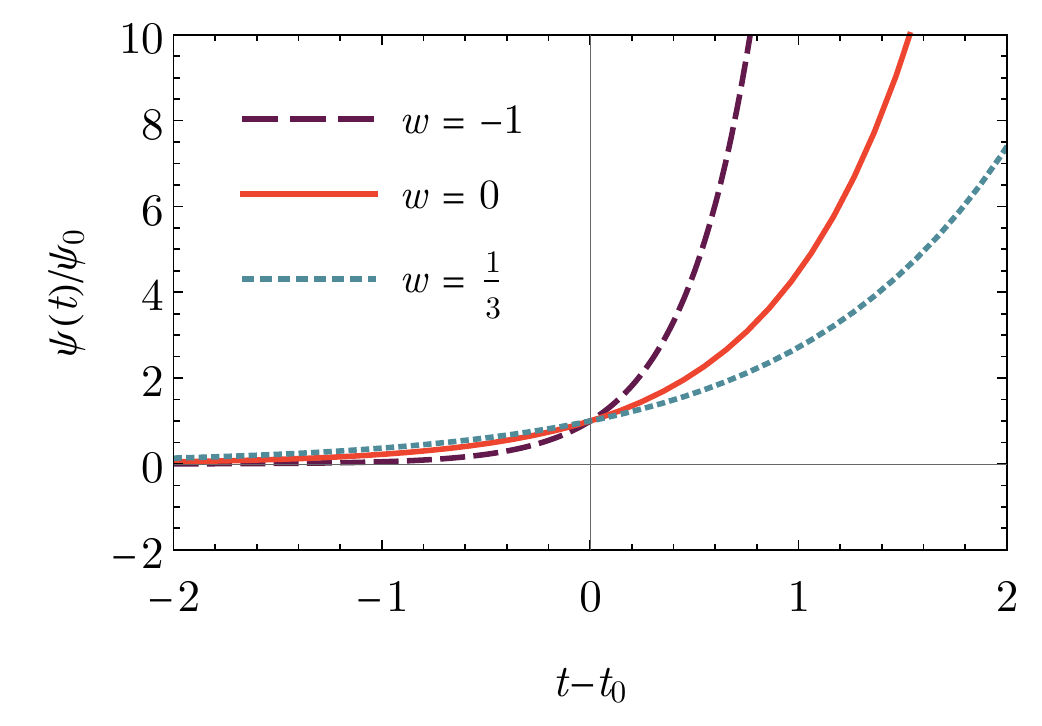}
	\caption{Scalar field $\psi\left(t\right)/\psi_0$ from Eq.~\eqref{eq:solpsi-dS} in the particular case where $a\left(t\right)$ is given by Eq.~\eqref{eq:a-dS}, setting $\Lambda=1$, with the equation of state forced to be each of these values $w=\{-1,0,1/3\}$.}
	\label{fig:Plot_dS_k0_psi}
\end{figure}

\begin{figure}
	\centering
	\includegraphics[width=0.6\columnwidth]{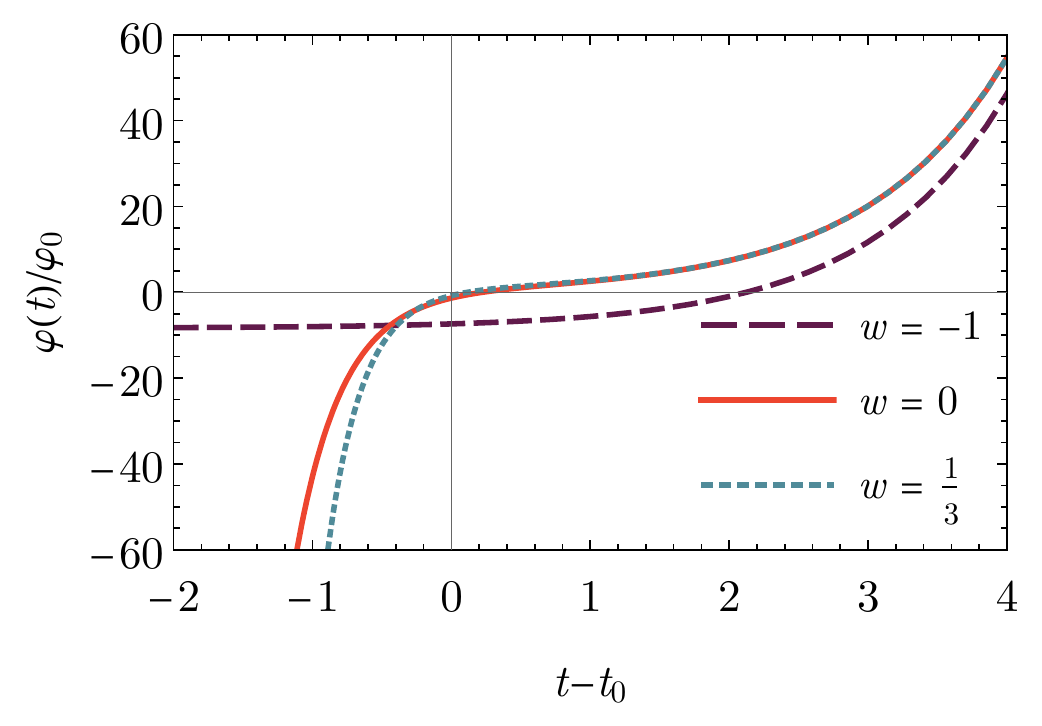}
	\caption{Scalar field $\varphi\left(t\right)/\varphi_0$ from Eq.~\eqref{eq:solphi-dSk0}, setting $\varphi_0=\rho_0=\psi_0=1$, in the particular case where $a\left(t\right)$ is given by Eq.~\eqref{eq:a-dS}, setting $\Lambda=1$, with the equation of state forced to be each of these values $w=\{-1,0,1/3\}$.}
	\label{fig:Plot_dS_k0_phi}
\end{figure}

For each of these three values of the equation of state, $w=\{-1,0,1/3\}$, the energy density as given by Eq.~\eqref{eq:solrho-dS} is plotted in Fig.~\ref{fig:Plot_dS_k0_rho}. When space expands exponentially, the energy density is constant for $w=-1$ (cosmological constant), and undergoes exponential decay for $w=0$ (dust) and $w=1/3$ (radiation) --- the decay is faster for radiation than for dust (due to the redshift of radiation). For each of the same three values of $w$, the scalar field $\psi(t)$ as given by Eq.~\eqref{eq:solpsi-dS} is plotted in Fig.~\ref{fig:Plot_dS_k0_psi}. In the three cases, the field $\psi$ grows exponentially with time --- in descending order, it grows faster with the cosmological constant than with dust and than with radiation. Likewise, we plot the scalar field $\varphi\left(t\right)$ as given by Eq.~\eqref{eq:solphi-dSk0} in Fig.~\ref{fig:Plot_dS_k0_phi}. When $t \gg t_0$, all of the plotted $\varphi\left(t\right)$ solutions grow exponentially (in fact, they can all be approximated by $\varphi\left(t\right) \sim \varphi_0 \exp(\sqrt{\Lambda}t)$ in this limit). On the other hand, when $t<t_0$ the $\varphi\left(t\right)$ solutions show different behaviour between them. The two scalar fields do not have a straightforward physical interpretation; as a reminder, they were defined as the partial derivatives of the function $f\left(R,T\right)$, i.e. $\varphi=\partial f / \partial R$ and $\psi=\partial f / \partial T$.
Finally, to each value of the equation of state corresponds a different form of the potential $V\left(\varphi,\psi\right)$ as given by Eq.~\eqref{eq:solV-dS-k0},
\begin{equation}\label{eq:solV-dSk0wneg}
w=-1: \hspace{0.5cm} V(\varphi,\psi)= V_0+12\Lambda\varphi-4\rho_0\psi,
\end{equation}
\begin{equation}\label{eq:solV-dSk0w0}
w=0: \hspace{0.5cm} V(\varphi,\psi)= V_0+12\Lambda\varphi+\frac{\rho_0\psi_0^2}{\psi},
\end{equation}
\begin{equation}\label{eq:solV-dSk0wpos}
w=1/3: \hspace{0.5cm} V(\varphi,\psi)= V_0+12\Lambda\varphi.
\end{equation}

 As exemplified here, we were able to find solutions for all the variables in our system: $\rho\left(t\right)$, $p\left(t\right)$, $\varphi\left(t\right)$, $\psi\left(t\right)$ and $V\left(\varphi,\psi\right)$, for particular choices of $a\left(t\right)$, $k$ and~$w$. The next question we addressed in our work was: what form(s) of the function $f\left(R,T\right)$ generate these cosmologies?

\subsection{Reconstructing the function $f\left(R,T\right)$}

Given that we were able to obtain consistent solutions, one may hope that it is possible to reconstruct the explicit form of the function $f\left(R,T\right)$ that can lead to those solutions. For instance, we have found the form of the potential $V\left(\varphi,\psi\right)$ and, by definition, the potential is related to the function $f\left(R,T\right)$, as per Eq.~\eqref{eq:potential} which can be rewritten as
\begin{equation}\label{eq:potentialRT}
f(R,T)=-V\left(f_R,f_T\right)+f_R R +f_T T,
\end{equation}
where we have changed the notation using $\varphi=f_R$ and $\psi=f_T$, with the subscripts $R$ and $T$ denoting the partial derivatives. Taking the partial derivatives of this equation with respect to $R$ and $T$, we obtained a system of two PDEs which are satisfied for an arbitrary $f\left(R,T\right)$ if $V_\varphi=R$ and $V_\psi=T$. These two conditions are the equations of motion we obtained for the fields $\varphi$ and $\psi$, in Eqs.~\eqref{eq:Vphi} and \eqref{eq:Vpsi}, and so are always satisfied. It can be seen, e.g. from Eq.~\eqref{eq:solV-dS-k0}, that 
\begin{equation}\label{eq:TfuncfT}
V_\psi =T= (3w-1)\rho_0\left(\frac{f_T}{\psi_0}\right)^{-\frac{2\left(1+w\right)}{\left(1-w\right)}},
\end{equation}
which is not defined for $w=1$. If $w=\{-1,1/3\}$, this equation does not give a relation between $f_T$ and $T$. On the other hand, if $w=0$, for instance, then $T=-\rho_0\psi_0^2 / f_T^2$. This relation can be inverted and integrated to give 
\begin{equation}\label{eq:fRT-general}
f(R,T)= g(R)\pm 2\psi_0\sqrt{-\rho_0 T},
\end{equation}
where $g(R)$ is an arbitrary function of $R$. In the case of a flat universe ($k=0$) with an exponentially evolving scale factor as given by Eq.~\eqref{eq:a-dS}, $V_\varphi=R=12\Lambda$ which does not provide a direct relation between $f_R$ and $R$, and so it does constrain the form of $g(R)$. Other particular cases are studied in our work and will be presented in the upcoming paper.

\section{Summary and concluding remarks}

Modified gravity has the potential to solve some of the remaining questions in cosmology, such as the current accelerated expansion. With this motivation, we took the recently developed scalar-tensor representation of the $f\left(R,T\right)$ gravity to explore allowed FLRW cosmologies. To keep our contribution short, we presented here only one particular case: that of an exponential expansion. As opposed to GR, in $f\left(R,T\right)$ gravity there is not a one-to-one relation between the evolution of the scale factor and the equation of state. For instance, we were able to find a solution by choosing a flat, dust dominated, exponentially expanding universe, which would not have been possible in GR. The extra degrees of freedom in modified gravity may contribute to an effective dark energy, thus allowing an exponential expansion to occur even when the equation of state is that of matter or radiation. In the paper under preparation, we include other particular cases, such as power law evolution of the scale factor, non flat geometries, and other equations of state. We are also already working on testing whether finite time singularities can appear in $f\left(R,T\right)$ gravity.

\section*{Acknowledgments}
	TBG is funded by a PhD Research Fellowship in the context of the Fundação para a Ciência e Tecnologia (FCT) project ``DarkRipple'' with reference PTDC/FIS-OUT/29048/2017.
	JLR was supported by the European Regional Development Fund and the programme Mobilitas Pluss (MOBJD647).
	FSNL acknowledges support from the Funda\c{c}\~{a}o para a Ci\^{e}ncia e a Tecnologia (FCT) Scientific Employment Stimulus contract with reference CEECINST/00032/2018, and funding from the research grants No. UIDB/FIS/04434/2020, UIDP/FIS/04434/2020, No. PTDC/FIS-OUT/29048/2017 and No. CERN/FIS-PAR/0037/2019.

\bibliographystyle{ws-procs961x669}


\end{document}